\documentclass[12pt,epsf]{iopart}
\usepackage{pstricks}
\usepackage{pst-coil}
\usepackage{amssymb}
\usepackage{graphics,epsf}
\begin{document}
\title{Dynamical extensions for shell-crossing singularities.}
\author{Brien C Nolan}
\address{School of Mathematical Sciences, Dublin City
University, Glasnevin, Dublin 9, Ireland.}
\eads{\mailto{brien.nolan@dcu.ie}}
\begin{abstract}We derive global weak solutions of Einstein's equations for
spherically symmetric dust-filled space-times which admit
shell-crossing singularities.  In the marginally bound case, the
solutions are weak solutions of a conservation law. In the
non-marginally bound case, the equations are solved in a
generalized sense involving metric functions of bounded variation.
The solutions are not unique to the future of the shell-crossing
singularity, which is replaced by a shock wave in the present
treatment; the metric is bounded but not continuous.
\end{abstract}
\submitto{\CQG}  \maketitle


\newcommand{\be}{\begin{equation}}
\newcommand{\ee}{\end{equation}}
\newtheorem{assume}{Assumption}
\newtheorem{theorem}{Theorem}
\newtheorem{prop}{Proposition}
\newtheorem{corr}{Corollary}
\newtheorem{lemma}{Lemma}
\newtheorem{definition}{Definition}
\newcommand{\me}{m^\epsilon}
\newcommand{\tm}{\tilde{m}}
\newcommand{\tme}{\tm^\epsilon}
\newcommand{\psie}{\psi^\epsilon}
\newcommand{\supp}{\hbox{supp}}
\newcommand{\gep}{G^\epsilon}
\section{Introduction}
The first examples of naked singularities occuring in
gravitational collapse which were constructed with a view to
testing and refining the cosmic censorship hypothesis were given
in \cite{yod1}. The singularities in question arise in the
collapse of an inhomogeneous dust sphere. One can consider the
dust sphere to be foliated by infinitely thin shells of matter;
the singularities occur when outer shells overtake inner ones, and
are known as shell-crossing singularities (SCS). These
singularities are not considered to be serious counter-examples to
the cosmic censorship hypothesis for resons which we review now
and will return to below.

The first and most convincing reason is that the fluid matter
model is not appropriate for the study of gravitational phenomenon
on the smallest scales. The fluid model is a macroscopic
approximation which works well when it can be interpreted as
representing the smoothed out behaviour of matter fields in a
region of space-time, as for instance in cosmology. Applying this
model all the way down to a scale where two infinitely thin shells
of fluid intersect is not appropriate. This view of the matter
model is now part of the cosmic censorship hypothesis, see e.g.\
\cite{wald}; the matter model being used must be such that it does
not develop singularities in special relativity. (It should be
pointed out that this view is not universally supported and that
some authors consider the shel-focussing singularity in fluid
models to be a valid counter-example to the cosmic censorship
hypothesis; see e.g.\ \cite{counter}.) In the case of the dust
matter model, which is a macroscopic model of collisionless
matter, this point of view has received independent support in
\cite{rendall}. It was shown here that the spherically symmetric
Einstein-Vlasov system, which involves a kinetic theoretic
description of collisionless matter and in which dust arises as a
singular limit, does not admit shell-crossing singularities. Note
that the occurrence of shell-crossing is not entirely due to the
vanishing pressure; an SCS can also occur when the pressure is
non-zero \cite{yod1}. It has also been shown that in spherically
symmetric perfect fluids, the global visibility of a
shell-crossing singularity is related to an unphysical situation,
namely the vanishing of the sound speed at the singularity
\cite{kriele}.

The second reason relates to the view that certain geometric
properties of shell-crossing singularities indicate the
possibility of extending the space-time beyond the singularity,
which would then not be considered genuine. While it is generally
accepted that geodesic incompleteness means that space-time is
singular, there does not exist a universally accepted definition
of what constitutes a genuine space-time singularity in the sense
of boundary constructions and levels of differentiability. However
an extension of even very low differentiability would allow one to
consider an SCS as an interior point of the space-time rather than
a boundary point (singularity). In particular, one could then test
Clarke's notion of generalized hyperbolicity for the SCS
\cite{clarke}. For example, it has been claimed that the metric
can be written in a form which is $C^0$ and non-degenerate at the
singularity \cite{newman} (but see below). Also in \cite{newman},
it was shown that the SCS is gravitationally weak in the sense of
Tipler: Jacobi fields carried along radial null geodesics running
into the singularity have finite limits at the singularity. It has
been claimed that this is evidence that an extension through the
singularity may be constructed. However not enough is currently
known about the connection between this definition of the
gravitational strength of a singularity and the question of
extendibility of the space-time to make this characterization
useful.

The third reason relates to actual attempts to construct an
extension. In \cite{papa}, a dynamical extension was constructed
for a (very) special case of an SCS. In \cite{clarke-o'd}, the
general case was treated. Motivated by behaviour observed in the
Newtonian case, the authors analysed the possibility of
constructing an extension through the singularity into a region
filled with three superimposed dust flows. The analysis is
incomplete, due to the awkward nature of the system of evolution
equations obtained, but strongly indicates the existence of a
solution to the extension problem. An extension along these lines
for the Newtonian case has recently been given in
\cite{clarke-swatton}. A model of shell-crossing involving the
collision of actual (distributional) shells of dust was studied in
\cite{klein-frauen}. The authors find that in order to single out
an extension to the future of the collision, one must put in extra
information by hand.

So the question remains open: does there exist a natural extension
through a shell-crossing singularity, and if so, is the extension
unique? We address these questions here and find affirmative and
negative answers respectively. The key is to introduce
co-ordinates which cast the field equations in a form in which
 the shell-crossing singularity is replaced by a shock
wave. We use $8\pi G = c =1$.

\section{Existence of weak solutions}
In co-moving co-ordinates, the line element for spherical
dust collapse is \cite{LTB}
\begin{equation}
ds^2 = -dt^2 +\frac{r_{,R}^2}{1+E}dR^2
+r^2(R,t)d\Omega^2,\label{eq1}
\end{equation}
where $d\Omega^2$ is the line element for the unit 2-sphere. The
field equations yield
\begin{eqnarray}
r_{,t}&=&-\sqrt{E+\frac{m}{r}},\label{feq1}\\
\rho &=& \frac{m^\prime(R)}{r^2r_{,R}},\label{feq2}
\end{eqnarray}
the latter equation defining the density $\rho(R,t)$ of the dust;
$m=m(R)$ and $E=E(R)$ are preserved by the fluid flow. These
functions are arbitrary (subject to certain energy and
differentiability conditions) and constitute the initial data for
the problem, set at $t=0$. $m$ is twice the Misner-Sharp mass. The
evolution equation (\ref{feq1}) is readily solved; in the general
case ($E\neq 0$) one obtains an implicit solution. From this,
$r_{,R}$ may be calculated. Regular initial data can lead to two
types of singularity: that occurring when $\rho$ diverges at $r=0$
(called the shell-focussing singularity), and that occurring when
$r_{,R}=0$ but $r\neq 0$. This latter is the shell-crossing
singularity. According to the implicit function theorem, the SCS
can be represented locally by a function $t=t_{SC}(R)>0$; this can
only be calculated explicitly in the marginally bound case $E=0$.
The time $t=t_{SF}(R)>0$ of the shell-focussing singularity can
always be calculated explicitly. See e.g.\ \cite{newman}. We say
that a shell-crossing singularity occurs if $t_{SC}(R)<t_{SF}(R)$
for some $R>0$; this indicates that the shell-crossing singularity
precedes the shell-focussing singularity. Note that if $t_{SC}$
and $t_{SF}$ are sufficiently smooth (in fact only continuous),
then the SCS occurs at values of $R$ forming disjoint open
subsets. Furthermore, it is possible to identify these open
subsets in terms of the initial data $E(R), m(R)$
\cite{hellaby+lake}. The co-ordinate freedom can be used to set
$r(R,0)=R$, and so the initial density is
$\mu:=\rho(R,0)=m^\prime/R^2$. For positive density in the matter
filled region, we require $m^\prime(R)>0$ for $R>0$, and so we see
that $\rho$ must diverge at the SCS. Note that if there exists a
region $t_{SC}<t<t_{SF}$, then the density given by (\ref{feq2})
is negative in this region. One needs a new form of the solution
which can be used beyond the SCS.

On one level, the SCS can be viewed simply as a breakdown of the
co-moving co-ordinate system: for fixed $t$, $R\mapsto r(R,t)$ is
no longer one-to-one. So we make the transformation
$(t,R)\to(t,r(R,t))$. This transformation was used in
\cite{newman}. The resulting line element is \[
ds^2=(1+E)^{-1}[-(1-\frac{m}{r})dt^2+2udtdr+dr^2] +r^2d\Omega^2,
\] where $u=\sqrt{E+m/r}$. This appears to yield a $C^0$ and non-degenerate metric at
the SCS. However this does not take account of the fact that $m,E$
are functions of $t$ and $r$. In particular, a space-time event on
the shell crossing corresponds to (at least) two values of the
co-moving co-ordinate $R$, and so one cannot uniquely determine
the value of $m(R)$ or $E(R)$ at the shell-crossing. Thus the line
element above does not provide an extension through the shell
crossing {\em unless} the functional dependence of $m$ and $E$ on
the proper time and proper radius is given; this principal aim of
this paper is to do this. The dependence in question is worked out
by solving the field equations in the new co-ordinate system.
These are simply the conditions that $m,E$ are preserved by the
fluid flow, and read
\begin{eqnarray}
m_{,t}-um_{,r}&=&0,\label{nfeq1}\\
E_{,t}-uE_{,r}&=&0.\label{nfeq2}
\end{eqnarray}

We define the characteristics of this system to be solutions of
\be \frac{dr}{dt}=-u.\label{char}\ee Then we see from
(\ref{nfeq1}) and (\ref{nfeq2}) that $m,E$ are constant along
characteristics and we can write down an implicit solution
obtained by integrating (\ref{char}). This solution holds until
such time as characteristics cross; the characteristics are the
fluid flow lines and so this is equivalent to shell-crossing. To
extend beyond the SCS, we let $f(u)=-u^2/2$ and rewrite the system
as
\begin{eqnarray}u_{,t}+f_{,r}&=&\frac{m}{2r^2},\label{lef1}\\
m_{,t}-um_{,r}&=&0.\label{lef2}
\end{eqnarray}
This system is fundamentally non-conservative, although
(\ref{lef1}) is an inhomogeneous conservation law which admits the
weak solution $u,m\in L^\infty(\mathbb{R}^2_+)$ if \be
\int_{\mathbb{R}^2_+}(u\psi_{,t}-\frac{u^2}{2}\psi_{,r}+\frac{m}{2r^2}\psi)\,drdt=0
\label{lef3}\ee for all test functions $\psi\in
C^\infty_0(\mathbb{R}^2_+)$. (\ref{lef3}) holds for differentiable
solutions of (\ref{lef1}) and extends the space of functions in
which we may seek solutions to $L^\infty(\mathbb{R}^2_+)$, where
$\mathbb{R}^2_+=(0,\infty)\times(0,\infty)$. In particular, a weak
solution need not be differentiable.
Noting that (\ref{lef1}), (\ref{lef2}) share the characteristic
speed $-u$, we must expect that discontinuities in $u,m$ will
occur at the same points. Then for (\ref{lef2}) we will need an
alternative solution concept to that used for the conservation law
(\ref{lef1}). We use an approach pioneered by Volpert
\cite{volpert} and developed by Le Floch \cite{lefloch}. In this
approach, we seek functions $u,m\in BV(\Omega)$, where
$\Omega=\mathbb{R}^2_+$ and $BV(\Omega)$ is the space of
real-valued functions of bounded variation on $\Omega$, i.e.\
functions locally integrable on $\Omega$ whose first order partial
derivatives are locally finite Borel measures \cite{volpert}. Such
functions have at worst jump discontinuities almost everywhere
(a.e.)\ on $\Omega$. If $u,v\in BV(\Omega)$ and (nonlinear) $g\in
C^1(\mathbb{R},\mathbb{R})$ then using a certain regularization
${\hat{g}}(u)$ of $g(u)$, the expression ${\hat{g}}(u)\partial
v/\partial x^i$ is a finite Borel measure which coincides with
$g(u)\partial v/\partial x^i$ a.e.\ on $\Omega$. See
\cite{volpert,lefloch} for details of these last two statements.
This allows us to make sense of the left hand side of (\ref{lef2})
when $u,m$ are both discontinuous. We define a weak solution of
(\ref{lef1}), (\ref{lef2}) to be a pair $u,m\in BV(\Omega)$ which
solve (\ref{lef1}) in the usual weak sense (i.e.\ which satisfy
(\ref{lef3}) for all $\psi\in C^\infty_0(\Omega)$) and which
satisfy $m_{,t}-\hat{u}m_{,r}=0$ in the sense of Borel measures.
The global existence of such solutions may be demonstrated as
follows. This extends the standard derivation of weak solutions
for conservation laws \cite{smoller}.

As noted above, $E,m$ are constant along characteristics and so a
classical solution is defined until such time as characteristics
intersect. Recall that shell-crossing (i.e.\ intersection of
characteristics) occurs in disjoint open subsets of the parameter
space labelled by $R$, which is the characteristic label. So we
may restrict attention to the case of one such open subset and
treat others in the same way. Let us first consider the nature of
the characteristics.

\begin{lemma}
Let $u_0, m_0\in C^1(\mathbb{R}_+)$ satisfy $u_0^\prime> 0$,
 $m_0^\prime\geq 0$, where $u_0(r)=u(r,0)>0$ and $m_0(r)=m(r,0)\geq 0$.
\begin{enumerate}
\renewcommand\labelenumi{\theenumi}
\renewcommand{\theenumi}{(\roman{enumi})}
\item The characteristic curves are $C^\infty$ on $\Omega$ and through
each point of $\Omega$, there passes at least one characteristic.
Each characteristic reaches $r=0$ in a finite amount of time.
\item If the characteristics $R_1,R_2$
(with $R_1<R_2$) intersect in the region $r>0$, then their slopes
at the point of intersection satisfy
\[ 0>\left.\frac{dt}{dr}\right|_{R_1}>\left.\frac{dt}{dr}\right|_{R_2}.\]
\item A given pair of characteristics intersects at most once and
does so at some time $t>0$.
\end{enumerate}
\end{lemma}

\noindent{\bf Proof:} Part (i) is proven by noting that
integrating (\ref{char}) is equivalent to integrating the
evolution equation (\ref{feq1}) in the co-moving co-ordinate
system. The characteristic $R$ (i.e.\ the solution of (\ref{char})
which satisfies $\left.r\right|_{t=0}=R$) reaches $r=0$ at the
shell focussing time $t_{SF}(R)$. To prove part (ii), let
$u_1,u_2$ denote the values of $u$ at the point of intersection on
the characteristics $R_1,R_2$ respectively. It suffices to show
that $u_2^2>u_1^2$. Let $r_3$ be the value of $r$ at which the
intersection occurs. Then $r_3<R_1<R_2$, since $R_1$ is the
initial value of $r$ along the characteristic $R_1$ and the
characteristics have negative slope. Then
\begin{eqnarray*}
u_2^2-u_1^2 &=& \frac{m_0(R_2)-m_0(R_1)}{r_3}+E_0(R_2) -
E_0(R_1)\\
&\geq&\frac{m_0(R_2)-m_0(R_1)}{R_2}+E_0(R_2) -
E_0(R_1)\\
&>&\frac{m_0(R_2)}{R_2}-\frac{m_0(R_1)}{R_1}+E_0(R_2) -
E_0(R_1)\\
&=&u_0^2(R_2)-u_0^2(R_1)>0\end{eqnarray*} as required. The proof
of part (iii) follows from the conclusion of part (ii); the
characteristic $R_1$ can only intersect $R_2$ from below.\hfill
$\square$

We note also that the regions in which characteristics intersect
form connected subsets of $\Omega$ which may be written explicitly
in terms of the characteristic label $R$ as
\begin{eqnarray*} Z=\{(r,t)\in\Omega: r=r(t;R),
t_{SC}(R)\leq t\leq t_{SF}(R), R_1\leq R\leq R_2\}\end{eqnarray*}
where $r=r(t;R)$ represents the characteristic with label $R$, and
$R_1,R_2$ are least and greatest values of $R$ for which the
corresponding characteristics meet others in $\Omega$.

Let $p$ be any point in $Z^o$, the interior of $Z$. We can
identify the left-most and the right-most characteristics which
intersect at $p$. Let these correspond to values $R_L<R_R$ of the
characteristic label respectively. There exists an open
neighbourhood $N_L$ of $R_L$ and a time interval $I$ such that the
change of coordinates $\gamma:(R,t)\mapsto (r(t;R),t)$ is a
diffeomorphism from $N_L\times I$ onto an open neighbourhood $N_p$
of $p$. (The Jacobean of this transformation is singular only at
the shell-crossing singularity, which corresponds to the boundary
of $Z$.) We can then define a function $u_L:N_p\to \mathbb{R}$ by
$u_L(r,t)=u_0(R)$, where $R$ is defined by
$(R,t)=\gamma^{-1}(r,t)$. From the assumptions and part (i) of
Lemma 1, we see that $u_L$ is $C^1$ on $N_p$. This extends to a
$C^1$ function on the interior of $Z$ and a corresponding $C^1$
map $u_R$ is defined in a similar fashion. Part (ii) of Lemma 1
implies that $u_L,u_R$ satisfy what we will refer to as a gap
property: $u_L(r,t)<u_R(r,t)$ for $(r,t)\in Z^o$.

Let $(r_0,t_0)$ be the co-ordinates of that point $p_0$ on the
shell-crossing singularity (i.e.\ on the boundary of $Z$) for
which $t_0$ is minimal. This is the globally earliest occurrence
of characteristic crossing, and we assume for convenience that
this defines a unique point. (This will involve an assumption
about the initial data $u_0,m_0$.) The existence, smoothness and
gap property of $u_L,u_R$ imply that there exist $C^0$ piecewise
$C^1$ curves $r=\phi(t)$ extending through $Z^o$ from $p_0$ to
$r=0$ with the property that
\[ -u_L(\phi(t),t)>\phi^\prime(t)>-u_R(\phi(t),t),\]
for all values of $t$ for which $\phi^\prime(t)$ exists. The
relevance of such curves is that they provide a means of
constructing a unique foliation of $\Omega$ by characteristics.
$\Omega-Z$ already admits such a foliation. In $Z$ and to the left
of $\Phi$ (the graph of $\phi$), we use the characteristics which
approach $\Phi$ from the left (those used to define $u_L$ along
$\Phi$). In $Z$ and to the right of $\Phi$, we use the
characteristics which approach $\Phi$ from the right. We note that
all uses of the words `right' and `left' are well-defined by the
gap property and the definition of $\phi$.

We are now in a position to write down weak solutions of
(\ref{lef1}) and (\ref{lef2}). For each path $\phi$ described
above, we define $m_\phi(r,0)=m_0(r)$ and then define $m_\phi$
throughout $\Omega-\Phi$ by taking $m_\phi$ to be constant along
the characteristics of the foliation constructed around $\phi$.
Taking $m_0\in C^1(\mathbb{R}_+)$, Lemma 1 shows that $m_\phi\in
C^1(\Omega-\Phi)$, with a jump discontinuity across $\Phi-p_0$.
This is sufficient to guarantee that $m_\phi\in BV(\Omega)$. We do
likewise to obtain $E_\phi\in BV(\Omega)$. This procedure yields
BV functions $m_\phi,E_\phi$ which solve the field equations
(\ref{nfeq1}) and (\ref{nfeq2}) along individual characteristics.
However more is required in order to have a weak solution. A
standard argument (see for example \cite{smoller}) shows that in
order for $u,m$ to satisfy (\ref{lef3}) for all $\psi\in
C^\infty_0(\Omega)$, the Rankine-Hugoniot condition must be
satisfied: \be [u]\dot\phi=[-\frac{u^2}{2}]. \label{rh}\ee For any
function $T$ on $\Omega$,  $[T]=T_+-T_-$ where $T_\pm$ denotes the
right and left-hand limits of $T(r,t)$ as the discontinuity is
approached along $t=$constant. Thus when $u_L\neq u_R$, i.e.\
anywhere in the interior of $Z$, we must have \be \dot\phi(t) =
-\frac12(u_L(\phi(t),t))+u_R(\phi(t),t)).\label{shockspeed}\ee The
smoothness of $u_L, u_R$ implies the existence and uniqueness of a
$C^2$ solution $r=\phi_*(t)$ of the initial value problem
consisting of (\ref{shockspeed}) and the initial condition
$r(t_0)=r_0$.

It remains to verify that (\ref{lef2}) is satisfied in the sense
of measures. This is trivially the case on $\Omega-\Phi_*$, where
the equation is solved in smooth functions. So consider the
measure $\sigma=m_{,t}-\hat{u}m_{,r}$. For any point $(r,t)$ on
the shock, this measure evaluates to \begin{eqnarray*}
\sigma(\{(r,t)\}) =
-[m]\dot{\phi}-[m]\int_0^1(u_L+\alpha(u_R-u_L))\,d\alpha =
0.\end{eqnarray*} We have used here Volpert's definition of the
regularization $\hat{u}$ of a discontinuous function, and applied
the Rankine-Hugoniot condition after evaluating the integral. This
concludes the proof of the existence of weak solutions of the
system (\ref{lef1}), (\ref{lef2}). We summarise as follows.

\begin{prop}
Let $u_0,m_0\in C^1(0,\infty)$ satisfy $u_0^\prime> 0$,
$m_0^\prime\geq0$. Then there exists $u,m\in BV(\Omega)$ giving a
weak solution of (\ref{lef1}), (\ref{lef2}) as defined above with
$u(r,0)=u_0(r), m(r,0)=m_0(r)$ and which satisfies $m(x+h,t)\geq
m(x,t)$ for all $h>0$ and all $(x,t)\in\Omega$.\hfill$\square$
\end{prop}

The last statement here is immediate from the construction and
indicates that the solutions have positive distributional density.
Note that the solutions are global in $t$, but that this does not
imply the absence of a future singularity. Figure One illustrates
the result of this Proposition for the initial data $E_0=0$ and
\be m_0(\xi)=\left\{ \begin{array}{ll}
  M(\xi), & 0\leq x \leq 2.25; \\
  M(2.25), & x> 2.25,
\end{array}\right.\label{masseg}\ee
where \[ M(\xi)= \frac{\xi^2}{(1+12\xi-9\xi^2+2\xi^3)^2}.\] Thus
the boundary of the dust sphere is initially at $x=2.25$.

\begin{figure}[one]
\centerline{\epsfxsize=10cm \epsfbox{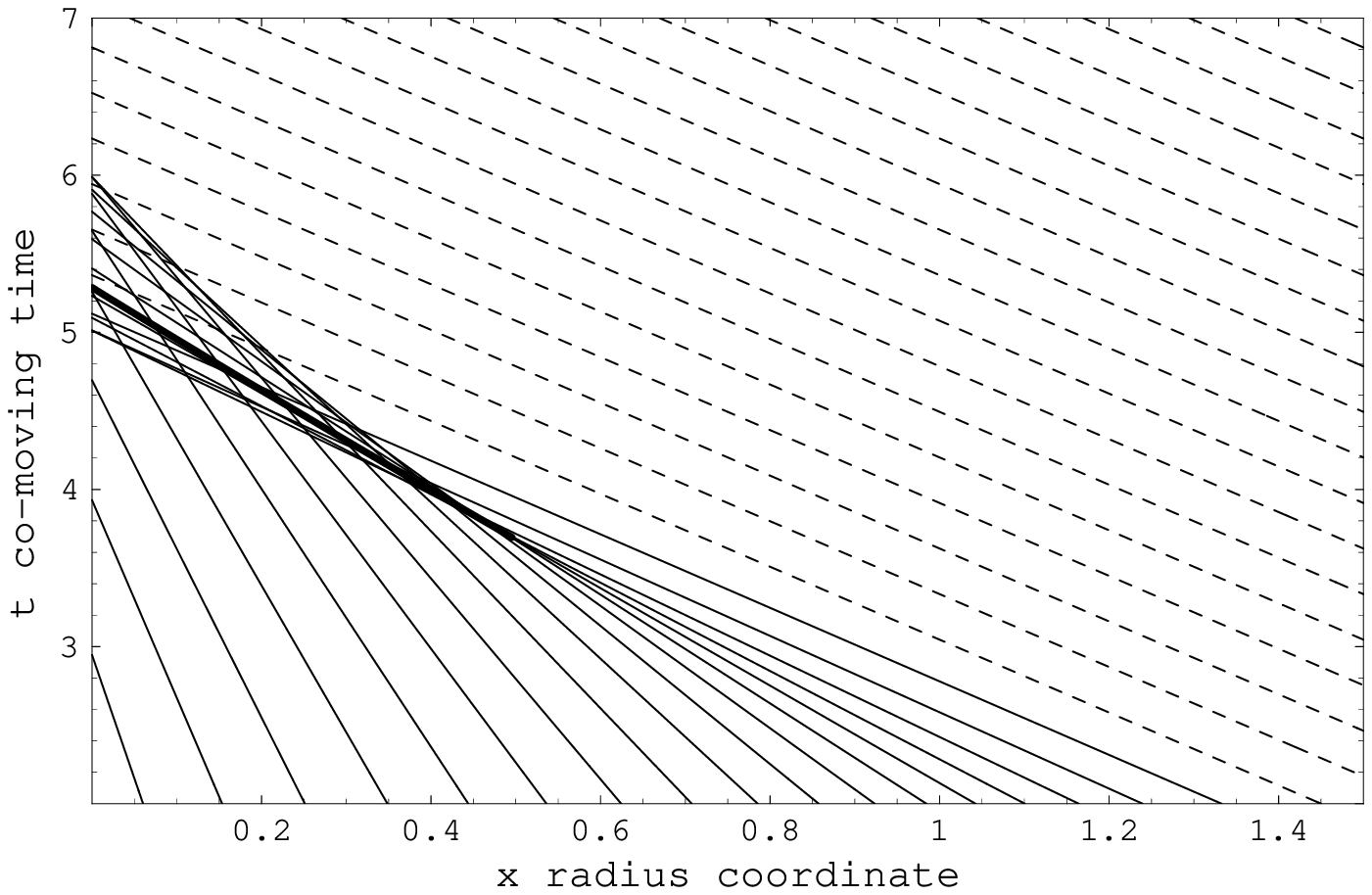}}
\caption{\label{Figure1} The diagram shows the characteristics
(solid and dashed lines) corresponding to the mass function
(\ref{masseg}). The dashed lines are the characteristics that
emanate from the vacuum region of the initial data surface $t=0$.
The shell-crossing singularity corresponds to the boundary of the
region $Z$ in which characteristics cross, the earliest point of
which is roughly at $(x,t)=(0.496,3.695)$. The shock, shown bold,
evolves from this point. (In the figure, this has been constructed
numerically using a second order Euler scheme.) The solution in
co-moving co-ordinates holds outside $Z$; the solution constructed
in Proposition One holds throughout $\Omega$ and so extends
through the shell-cross.}
\end{figure}

\section{Uniqueness considerations}
In both the marginally and non-marginally bound cases,
uniqueness fails on a fundamental level. To see this we focus on
the marginally bound case. In this case,  $E=0$ and there is only
field equation (\ref{nfeq1}) which can be written in conservative
form by taking $x=-\frac23r^{3/2}$; \[
m_{,t}-\sqrt{m}m_{,x}=m_{,t}+f_{,x}=0\] where $f=-\frac23m^{3/2}$.
For this form of the equation, the Rankine-Hugoniot condition is
$[m]\dot{\phi}=[f]$. For any smooth function $M(m)$, the solution
of the  pde's for $M$ and $m$ agree in the smooth domain. However
the conservative form of the equation for $M$ {\em will not} be
the same as that for $m$. This leads to a different shock speed
(\ref{shockspeed}), and hence a different location for the shock.
For example, $M=\ln{m}$ leads to $[\ln
m]{\dot{\phi}}=[-2\sqrt{m}]$. Thus we do not have a unique weak
solution for given initial data; we could only expect to obtain
such if the field equations were fundamentally integral
conservation laws which is not the case.

However it is possible to obtain a uniqueness result on a
different level for the marginally bound case. (There are no
global uniqueness results for weak solutions of non-conservative
first order systems.) We note that as with entropy solutions, the
condition which yields uniqueness has a strong physical
motivation, being equivalent  in the present case to positivity of
the energy density considered as a measure. The proof involves a
variation of Oleinik's proof of the uniqueness of entropy
solutions of conservation laws with convex flux functions (see
chapter 16 of \cite{smoller}).

\begin{prop}
Let $m_0\in L^\infty(\mathbb{R}_+)$ satisfy $m_0(x+h)\geq m_0(x)$
for all $x\geq0$ and $h>0$. Let $f(m)=-\frac23 m^{3/2}$. Then
there exists a unique weak solution on $\mathbb{R}^2_+$ of the
conservation law \be m_{,t}+f(m)_{,x}=0\label{sfeq}\ee which
satisfies
\[ m(x+h,t)\geq m(x,t) \]
for all $h>0$ and for all $(x,t)\in \mathbb{R}^2_+$, and
$m(x,0)=m_0(x)$.
\end{prop}

{\bf Proof:} We refer to a solution described in this proposition
as a positive density solution.

(i) As regards existence, the proof of Proposition 1 is easily
adapted to the present case. The regularity of the data can be
lowered by virtue of the fact that the characteristics in this
case are straight lines satisfying \[ \frac{dx}{dt}=-\sqrt{m}.\]

(ii) For uniqueness, our aim is to prove that there is at most one
positive density solution in $L^\infty(\Omega)$. So let $m,\tm$ be
two such solutions with the same initial data. Then each satisfies
the weak solution condition
\[ \int_\Omega (m\psi_{,t}+f(m)\psi_{,x})\,dxdt=0\]
for all $\psi\in C^\infty_0(\Omega)$, and so we can write
\begin{eqnarray} 0&=&\int_\Omega
[(m-\tm)\psi_{,t}+(f(m)-f(\tm))\psi_{,x}]\,dxdt\nonumber\\
&=&\int_\Omega
w[\psi_{,t}+G(m(x,t),\tm(x,t))\psi_{,x}]\,dxdt,\label{eq2}\end{eqnarray}
where $w=m-\tm$ and
\[ G=\int_0^1 f^\prime(\lambda
m(x,t)+(1-\lambda)\tm(x,t))\,d\lambda.\] The proof proceeds by
attempting to replace the term multiplying $w$ in the integrand of
(\ref{eq2}) by an arbitrary test function $\phi\in
C^\infty_0(\Omega)$. The result would then be immediate, but the
non-smoothness of $G$ must be dealt with carefully. As mentioned
above, this proof follows almost identically Oleinik's proof of
uniqueness of entropy solutions as given in \cite{smoller}.

(iii) For any $\epsilon\in(0,1)$, we define the smooth functions
$\me=\eta_\epsilon\ast m$ and $\tme=\eta_\epsilon\ast \tm$ where
$\eta_\epsilon$ is the standard mollifier and $\ast$ represents
convolution. Then $\|\me\|_\infty\leq\|m\|_\infty$ and $\me\to m$
pointwise a.e. in $\Omega$, with similar results holding for
$\tme$. The positive density condition ($m_0$ non-decreasing)
implies that \be \me_{,x}(x,t)\geq 0 \label{pd}\ee in $\Omega$.

Next, we define $\gep = G(\me,\tme)$ and write (\ref{eq2}) as
\begin{eqnarray}
0&=&\int_\Omega w(\psi_{,t}+\gep\psi_{,x})\,dxdt+\int_\Omega
w(G-\gep)\psi_{,x}\,dxdt.\label{eq3}
\end{eqnarray}
Now let $\phi\in C^\infty_0(\Omega)$ be arbitrary, and take $T>0$
such that $\supp(\phi)\subseteq (0,\infty)\times(0,T)$. Let
$\psie$ be the solution of the terminal value problem
\begin{eqnarray}
\psie_{,t}+\gep\psie_{,x}&=&\phi \quad \hbox{in} \quad
(0,\infty)\times(0,T)\nonumber\\
\psie&=&0\quad \hbox{on}\quad (0,\infty)\times\{t=T\}.
\label{eq4}\end{eqnarray} This problem may be solved by the method
of characteristics and admits a unique smooth solution.
Furthermore, since $|\gep|$ is bounded in the compact set
$\supp(\phi)\subset\Omega$, the solution $\psie$ also has compact
support in $(0,\infty)\times[0,T)$.

(iv) We next establish the following upper bound: \be
|\psie_{,x}(x,t)|\leq e^T\|\phi_{,x}\|_\infty \quad \hbox{in}
\quad \supp(\psie).\label{bound}\ee

Note that
\[ \gep_{,x}=\int_0^1
f^{\prime\prime}(\lambda\me+(1-\lambda)\tme)(\lambda\me_{,x}+(1-\lambda)\tme_{,x})\,d\lambda.\]
Since $m>0$ in $\Omega$, $f^{\prime\prime}=-\frac12m^{-1/2}$ is
negative and bounded on any compact subset of $\Omega$. The
positive density condition (\ref{pd}) then yields $\gep_{,x}\leq0$
in $\supp(\psie)$.

Let $a=e^t\psie_{,x}$. From (\ref{eq4}), we obtain \be a_{,t}+\gep
a_{,x}=(1-\gep_{,x})a+e^t\phi_{,x},\label{eq5}\ee and we note that
$a=0$ on $t=T$. Let $s>0$. In the compact set
$K_s=\supp(\psie)\cap\{s\leq t\leq T\}$, $a$ attains its maximum
and minimum values. Let its maximum occur at $(x_0,t_0)$; we must
have $a(x_0,t_0)\geq 0$. If $(x_0,t_0)\in \partial K_s$, with
$t_0>s$, then $a(x_0,t_0)=0$. If $s\leq t_0<T$ and $(x_0,t_0)$ is
in the interior of $K_s$, then we must have
\[ a_{,t}\leq0,\quad a_{,x}=0\]
at the maximum. Then from (\ref{eq5}), at $(x_0,t_0)$ we have
\begin{eqnarray*}
0&\geq& (1-\gep_{,x})a+e^{t_0}\phi_{,x}\\
&\geq&a+e^{t_0}\phi_{,x}.
\end{eqnarray*}
From this we obtain \begin{eqnarray*}
a_{max}&\leq&-e^{t_0}\phi_{,x}\\&\leq&
e^{t_0}\|\phi_{,x}\|_\infty\\
&\leq& e^T\|\phi_{,x}\|_\infty.
\end{eqnarray*}

Similarly, at the non-positive minimum we must have either $a=0$
or
\[ a_{,t}\geq 0,\quad a_{,x}=0.\] In this case, (\ref{eq5}) yields
\begin{eqnarray*}
0&\leq& (1-\gep_{,x})a+e^{t_0}\phi_{,x}\\
&\leq&a+e^{t_0}\phi_{,x}.
\end{eqnarray*}
From this we obtain \begin{eqnarray*}
a_{min}&\geq&-e^{t_0}\phi_{,x}\\
&\geq& -e^T\phi_{,x}\\
&\geq& -e^T\|\phi_{,x}\|_\infty.
\end{eqnarray*}

Hence
\[ |a|\leq e^T\|\phi_{,x}\|_\infty,\]
and so we obtain the $s-$independent bound \[ |\psie_{,x}|\leq
e^T\|\phi_{,x}\|_\infty \] on $\supp(\psie)\cap \{s\leq t\leq T\}$
for any $s>0$, and so we may let $s\to 0$ to complete the
derivation of (\ref{bound}).

(v) It remains to show that for any $\phi\in C^\infty_0(\Omega)$
and $\delta>0$,
\[ \left| \int_\Omega w\phi\,dxdt\right|<\delta.\]
In (\ref{eq3}), take $\psi$ to be $\psie$, the solution of
(\ref{eq4}). Then
\[ \int_\Omega w\phi\,dxdt = -\int_\Omega
w(G-\gep)\psie_{,x}\,dxdt.\] The domain of integration on the
right hand side may be replaced by $\supp(\psie)$, whereon
(\ref{bound}) applies to bound $|\psie_{,x}|$ by a constant. Then
the pointwise limit $G-\gep\to0$ yields the result by taking
$\epsilon$ to be sufficiently small.\hfill$\square$

We note that this proof carries through for different formulations
of the field equation (\ref{sfeq}). That is, for a diffeomorphism
$G:\mathbb{R}^+\to\mathbb{R}^+$ with $G(0)=0$, the corresponding
equation for $M=G(m)$ with data corresponding to positive density
has unique weak positive density solutions. These solutions will
however disagree with that of Propostion 2.

\section{Discussion}

The solutions presented here show that one can extend space-time
beyond a shell-crossing singularity in a natural way; the
extension arises almost immediately when one steps out of the
co-moving co-ordinate system. The drawback is that the extension
is not unique. Our interpretation of this fact is that, leaving
aside the question of the matter model, shell-crossing
singularities should be included in a discussion of cosmic
censorship. This is based on the following argument originally put
forward in \cite{moncrief-eardley}. Strong cosmic censorship
asserts that generically, initial data evolve to give a maximal
globally hyperbolic space-time. It has been shown that such an
evolution must be unique \cite{cb-geroch}. Thus the lack of a
unique evolution indicates a violation of strong cosmic
censorship. Genericity aside, this is the situation we have found
here: there are multiple $L^\infty$ weak solutions of Einstein's
equation for spherical dust for certain open subsets ({\em cf.}
\cite{hellaby+lake}) of the space of initial data.

Having extended beyond the shell-crossing singularity, one can now
investigate the question of the global structure of the space-time
and generalized hyperbolicity and study geodesics impinging on the
singularity. This work is in progress, utilizing Colombeau theory
\cite{colombeau} as has recently been done for space-times
admitting conical singularities \cite{vickers-wilson},
\cite{wilson} and for hypersurface singularities
\cite{vickers-wilson2}.

The problem of uniqueness seems hard to resolve, as it would seem
to involve deciding on which formulation of the conservation law
(\ref{lef1}) is `correct'. One possible way to obtain uniquely the
shock surface is to consider the conservation law
$\nabla_aT^{ab}=0$. This was the route taken in
\cite{smoller-temple}, in which the authors generalize the classic
results of Israel \cite{israel} to Lipschitz continuous metrics
and so deal with a different variety of gravitational shocks. In
our the present case we do not have Lipschitz continuity. Indeed
$T^{ab}$ is measure-valued and it is far from clear how to extract
jump conditions from the conservation law. An approach based on
the recently developed geometric theory of nonlinear generalized
functions \cite{diana} may yield the required results.

There is an interesting analogy between shell-crossing
singularities and shock waves in gas dynamics. In each case, the
singularity disappears when a more appropriate matter model is
used (dust replaced by the Vlasov model, inviscid fluid replaced
by viscous). In the latter case, the shocks observed in the
inviscid system reflect large derivatives which can arise in the
more realistic model. It would be of interest to see if this
pattern is repeated in the dust-Vlasov pair, where the existence
of large derivatives in the latter model may be to all intents and
purposes physical singularities.

\ack I am grateful to G. H\"ormann, M. Kunzinger, M.
Oberguggenberger and R. Steinbauer for enlightening discussions
and to the Royal Irish Academy and the Austrian Academy of
Sciences for financial support. I also thank C. Barrab\`{e}s and
P. Hogan for their support and the referees for their comments.

\end{document}